\documentclass[]{pasj01}
\draft

\begin{document} 
\Received{}
\Accepted{}

\title{A Search for a keV Signature of Radiatively Decaying Dark Matter with Suzaku XIS Observations of the X-ray Diffuse Background}

\author{Norio \textsc{sekiya}, Noriko Y. \textsc{yamasaki}, and Kazuhisa \textsc{mitsuda}}%
\altaffiltext{}{Institute of Space and Astronautical Science, Japan Aerospace Exploration Agency,\\
3-1-1 Yoshinodai, Sagamihara, Chuo-ku, Kanagawa 252-5210, Japan}
\email{yamasaki@astro.isas.jaxa.jp}


\maketitle

\begin{abstract}
We performed the deepest search for an X-ray emission line between 0.5 and 7 keV from non-baryonic dark matter with the Suzaku XIS.
Dark matter associated with the Milky Way galaxy was selected as the target to obtain the best signal-to-noise ratio.
From the Suzaku archive, we selected 187 data sets of blank sky regions which were dominated by the X-ray diffuse background. The data sets were from 2005 to 2013.
Instrumental responses were adjusted by multiple calibration data sets of the Crab Nebula.
We also improved the technique of subtracting lines of instrumental origin.
These energy spectra were well described by X-ray emission due to charge exchange around the Solar System, hot plasma in and around the Milky Way and superposition of extra-galactic point sources.
A signal of a narrow emission line was searched for, and the significance of detection was evaluated in consideration of the blind search method (the Look-elsewhere Effect).
Our results exhibited no significant detection of an emission line feature from dark matter.
The 3$\sigma$ upper limit for the emission line intensity between 1 and 7 keV was $\sim10^{-2}$ photons cm$^{-2}$ s$^{-1}$ sr$^{-1}$, or $\sim 5\times10^{-4}$ photons cm$^{-2}$ s$^{-1}$ sr$^{-1}$ per $M_\odot$ pc$^{-2}$, assuming a dark matter distribution with the Galactic rotation curve.
The parameters of sterile neutrinos as candidates of dark matter were also constrained.
\end{abstract}

\section{Introduction}
The existence of ``dark matter'' (DM) in the Universe has been demonstrated by modern astrophysical and cosmological observations, such as rotation curves of spiral galaxies, masses of intra-cluster gas, gravitational lensing of clusters of galaxies, and observational data of the Cosmic Microwave Background anisotropy and Large Scale Structure.
It is considered to constitute about a quarter of the total energy density in the present Universe and to occupy more than 80 \% of the total mass density.
The current ``standard cosmology'', the $\Lambda$-Cold Dark Matter ($\Lambda$CDM) model, assumes that dark energy and DM play prominent roles in gravitational effects in the formation of structure in the Universe.
Since nucleosynthesis models of the early Universe place limits on the fraction of baryons, DM cannot be baryonic matter, which forms a part of the Standard Model (SM) of particle physics.
The nature of DM is thus an issue of fundamental significance.
Several candidates such as ``WIMPs'' (weakly interacting massive particles), ``super-WIMPs'' and ``sterile neutrinos'' postulated, and either direct or indirect DM searches have been conducted to find them.
However, none of them have been successful at present.
Searching for photon emission from the decay or annihilation of DM particles through astrophysical observations is a promising approach to the discovery of DM, and the X-ray region is one of the possible windows.
In particular, sterile neutrinos as DM candidates attract a lot of attention \citep{2001ApJ...562..593A,2002APh....16..339D,2012PDU.....1..136B}.
They could be generated sufficiently in the early Universe through given mechanisms, for example the non-resonant or resonant production \citep{1994PhRvL..72...17D,1999PhRvL..82.2832S} and decay of scalar field (e.g. \cite{2006PhRvL..97x1301K,2006PhLB..639..414S,2008PhRvD..77f5014P}).
The relic sterile neutrino abundance from scattering-induced conversion of active neutrinos was first analytically estimated by \citet{1994PhRvL..72...17D} and able to account for all of DM.
The model containing sterile neutrinos is strongly motivated by the neutrino flavor oscillation which is supported by the atmospheric neutrino evidence of the Super-Kamiokande \citep{1998PhRvL..81.1562F}, the baryon asymmetry of the Universe and other curious things beyond the SM \citep{2005PhLB..620...17A,2005PhLB..631..151A}.
The flavor oscillation between sterile neutrinos and active neutrinos or radiative decay is predicted although its mixing angle may be very small.
Associated with the decay, a photon with the energy $E = m_{\rm s}c^2/2$ are emitted ($m_{\rm s}$ is the mass of sterile neutrinos, $c$ is the speed of light).
A keV-mass sterile neutrino is a Warm DM (WDM) candidate \citep{2001PhRvD..64b3501A}.
It resolves several inconsistencies between the predictions of the CDM model and the observational results such as the shape and smoothness of DM halos \citep{2006MNRAS.368.1073G,2007ApJ...663..948G,2008IAUS..244...44W,2014MNRAS.439..300L}.
Since the keV-mass sterile neutrinos should decay and produce a keV X-ray photon, a search for this radiative decay emission line in the X-ray range is meaningful.
Radiatively decaying DM such as sterile neutrinos have been vigorously searched for by X-ray observatories.
\citet{2007A&A...471...51B}, \citet{2014PhRvL.113y1301B}, \citet{2014PhRvD..89b5017H} and \citet{2014PhRvD..90j3506M} searched DM in the Local Group, i.e. the Milky Way galaxy, its satellite dwarf galaxies and M31, and obtained the tightest restriction on the DM line intensity.
In 2014, several papers reported a possible X-ray emission line around 3.5 keV.
In \citet{2014ApJ...789...13B}, the first report of this line, XMM-Newton observational spectra of 73 clusters of galaxies were used to search for a non-baryonic emission line.
The significance of this detection was as high as 4.3$\sigma$.
Independently, \citet{2014PhRvL.113y1301B} found this line in the Perseus cluster and M31 with the 4.4$\sigma$ significance.
Other groups \citep{2015PASJ...67...23T,2015JCAP...02..009C}, however, could not confirm this detection with the same instruments or the same targets and gave the upper limits.
At the center of the Milky Way galaxy, \citet{2014arXiv1408.2503} reported to detect the 3.54 keV line by XMM-Newton, while \citet{2014arXiv1405.7943R} and \citet{2015MNRAS.450.2143J} obtained upper limits by Chandra and by XMM-Newton, respectively.
Due to the complexity of the instrument responses and possible astrophysical emission (e.g. K\emissiontype{XVIII} around 3.5 keV), a search for DM requires careful studies.
In this paper, we selected the best observational target for the DM search in the keV energy range with X-ray CCD instruments and searched for a non-baryonic X-ray signature to provide valuable constraints on the parameter space of DM.
All error ranges in our text and tables correspond to 90 \% confidence levels, and vertical error bars in our figures indicate 1$\sigma$.
Throughout this paper, we assume the following values for cosmological parameters: the energy density parameters $\Omega_{\rm m}=0.27$, $\Omega_\Lambda=0.73$ and the Hubble constant $h_0=0.7$.
We utilized Ftools in HEAsoft version 6.15 and XSPEC version 12.8.1\footnote[1]{http://heasarc.gsfc.nasa.gov/lheasoft/} for this analysis.
\section{Strategy of the analysis}
\subsection{Selection of targets and instruments}\label{Section2.1}
Selecting suitable targets and instruments is important in searching for DM in the keV energy range.
Gravitational sources with high DM column density such as galaxies and clusters of galaxies are plausible candidate targets.
Since we are located within the DM distribution of the Milky Way, we also have the potential to detect its signal over the whole sky, even in the blank sky.
Such targets, however, contain high temperature plasma and so are sources of X-rays as background emission.
The DM column density in the Perseus cluster or a rich cluster of galaxies was estimated as $\sim 800$ $M_\odot$ pc$^{-2}$  \citep{2014ApJ...789...13B,2015PASJ...67...23T}.
That of M31 is $\sim 600$ (200 -- 1000) $M_\odot$ pc$^{-2}$ (see \cite{2014PhRvL.113y1301B} and references therein).
We estimated the distribution of the DM within the Milky Way by the NFW profile with parameters from \citet{2012PASJ...64...75S}, as $\rho_0=1.06\times 10^{-2}$ $M_\odot$ pc$^{-2}$ and $h=12.5$ kpc, and calculated the column density as a function of an angle from the Galactic center (GC).
It is $\sim 200$ $M_\odot$ pc$^{-2}$ towards the GC and decreases to 40 $M_\odot$ pc$^{-2}$ at the anti-Center.
The sensitivity of the line detection is limited by the background emission.
We simulated mock Suzaku spectra for the Perseus cluster, M31, and a typical blank sky observation with the same field of view (FoV) and exposure time.
The blank sky is occupied by the X-ray diffuse Background (XDB) which consists of plasma emission in the Milky Way and a superposition of extragalactic point sources.
Typically, the surface brightness of the Perseus or M31 are $\sim 10000$ or $\sim 100$ times brighter, and the line detection limits for them are $\sim 100$ or $\sim 10$ times higher than that for the blank sky.
The meaning of ``typical'' is that there are much line emission from the associated plasma, which hinder the DM detection at various energies even if they are modeled well.
Potential sensitivity for the DM detection is evaluated by the DM column density divided by the line detection limit.
The sensitivity for M31 and the blank sky are almost at the same level, which is $\sim 10$ times better than that for the Perseus cluster.
We then evaluated the sensitivity based on the actual observation cases by X-ray observatories, Suzaku, XMM-Newton and Chandra, with realistic instrumental responses and Non-X-ray Background (NXB), as well as available exposure time and sky coverage of archival data.
The total exposure time for archival data of the Perseus cluster is larger than that of M31, but could not compensate the sensitivity limited by the cluster emission as background.
In the archives, there are hundreds of observational data sets of blank sky regions or maskable faint compact sources which are dominated by the XDB and used to search for DM associated with the Milky Way.
The total photon statistics for these XDB data is larger than that for the other targets including M31.
Due to the low surface brightness of the XDB, however, the NXB would also limit the sensitivity.
The NXB of XMM-Newton (MOS, PN) and Chandra (ACIS-I, ACIS-S) for corresponding sky coverage are comparable to surface brightness levels between that of M31 and XDB observations.
Thus the line sensitivity for the XDB observation by XMM-Newton and Chandra are limited by their NXB.
On the other hand, the X-ray Telescope and X-ray Imaging Spectrometer of Suzaku (XRT-XIS; \cite{2007PASJ...59S...1M,2007PASJ...59S...9S,2007PASJ...59S..23K}) combined to give the lowest and most-stable NXB owing to combination of the low-Earth orbit and the instrumental design.
Using Suzaku XIS observational data of the XDB is therefore the most suitable for this study.
We note that we gave up to use the GC regions because there are number of bright point sources and diffuse structure which are difficult to be modeled and subtracted.
In this paper, we used cleaned event files of the FI-CCDs (XIS0, 2 and 3) and the BI-CCD (XIS1) of Suzaku.
\subsection{Archival data selection}
\begin{figure}[htbp!]
\begin{center}
   \includegraphics[width=\linewidth]{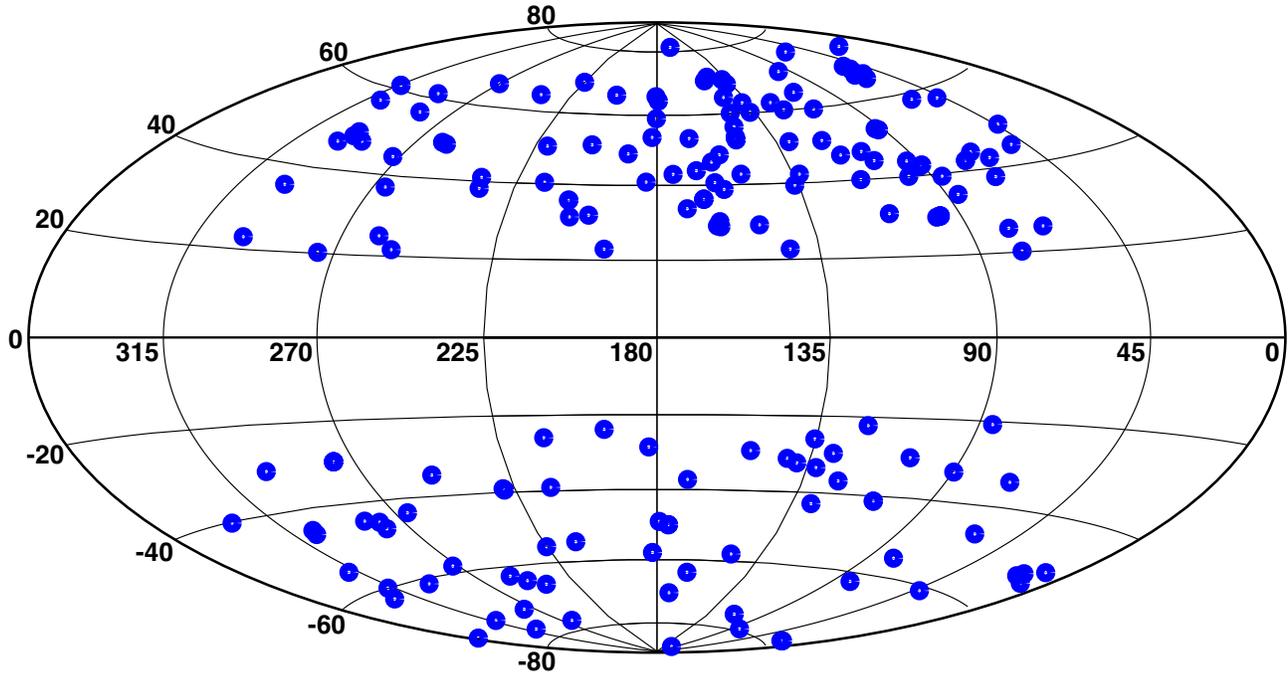} 
\caption{187 regions used to search for a keV signature of DM from Suzaku archival data.
These are plotted on the all sky map with the Galactic coordinate system centered at the Galactic anti-center.
}
\label{Figure1}
\end{center}
\end{figure}

We collected approximately-pure XDB data in the Suzaku XIS archive from 2005 to 2013 that satisfied the requirements shown below:
\begin{enumerate}
\item No bright sources (i.e. original observational purpose are blank sky fields study) or maskable faint compact sources are in the XIS FoV.
\item Galactic latitudes $|b|>20^\circ$ to avoid the X-ray emission peculiar to the Galactic disk \citep{2009PASJ...61S.115M}.
\item Separate from the region occupied by the North Polar Spur and other local X-ray sources.
\item The XIS was operated in the normal clocking mode and the $3\times 3$ or $5\times 5$ editing mode.
\end{enumerate}
Ultimately, 187 Suzaku XIS observational data sets were selected as shown in Fig. \ref{Figure1}.
We collected more XDB data than has been observed previously by the Suzaku XIS.
Since many complex unresolvable emission appears in the low energy range ($<0.7$ keV) and XDB photons are not adequately available (NXB photons dominate) in the high energy range ($>5$ keV), we used the 0.7 -- 7 keV and the 0.5 -- 5 keV ranges for the FI-CCDs (XIS0, 2 and 3) and the BI-CCD (XIS1), respectively.
Signals from anomalous (hot and flickering) pixels in the XISs were screened out with the Ftool {\tt cleansis}.
\section{Data reduction and reproduction of instrumental responses}
\subsection{Data reduction and point source removal}
We first conducted an imaging analysis over all of the data in order to reject resolvable X-ray point sources contaminating the XDB spectra.
The procedure was the following:
\begin{enumerate}
\item XIS images of the 0.5 -- 7 keV range were extracted.
\item Point sources in the XIS FoVs were detected and rejected by using a wavelet function of similar size to the point spread function of the XRT-XIS using {\tt wavdetect} from the Chandra Interactive Analysis of Observations, version 4.6.
\item Point sources whose flux was larger than $1.0\times 10^{-14}$ erg cm$^{-2}$ s$^{-1}$ in the 0.5 -- 7 keV range were removed with circular regions centered at their positions.
The radius of the circular regions $>1.5'$ were determined such that these regions included $>90$ \% of source photons.
\end{enumerate}
Second, in order to remove X-ray emission from the Earth's atmosphere and from the Solar wind, and to reduce the NXB, we also selected good time intervals by applying the following criteria, as was done in \citet{2014PASJ...66L...3S}:
\begin{enumerate}
\item The elevation angle from the bright/dark Earth limb was chosen as $>20^\circ$/$5^\circ$ to avoid fluorescent emission lines from the Earth's atmosphere.
\item Time intervals excluding the South Atlantic Anomaly passage 
\item The Cut Off Rigidity (COR2) $>8$ GV $c^{-1}$ to reduce the high-energy-charged-particle background due to the low Earth's magnetic field \citep{2008PASJ...60S..11T}.
\item Time periods when the proton flux in the Solar wind fell below the typical threshold, $4.0\times 10^8$ cm$^{-2}$ s$^{-1}$, to lower effects of the Geocoronal-Solar wind charge exchange \citep{2007PASJ...59S.133F}.
The proton flux was observed with monitoring satellites: ACE/SWEPAM\footnote[2]{http://www.srl.caltech.edu/ACE/ASC/level2/lvl2DATA\_SWEPAM.html} and WIND/SWE\footnote[3]{http://web.mit.edu/space/www/wind\_data.html}.
\end{enumerate}
After this data screening, the total exposure time becomes 31.5 Msec $\sim 1$ year, and we obtained $\sim 2\times 10^6$ counts of photons in the 0.5 -- 7 keV range.
This is the largest data for the XDB with the Suzaku XIS.
\subsection{Nominal instrumental responses and NXB reproduction}
We then reproduced instrumental responses at the time of each observation.
The energy redistribution matrix files (RMFs) including quantum efficiency and energy responses (energy scale and resolution) of the XISs were generated by the Ftools {\tt xisrmfgen}.
The ancillary response files (ARFs) involving angular responses and effective area of the XRT-XIS modules were produced by the Suzaku calibration database and Monte-Carlo ray-tracing simulations: the Ftools {\tt xissimarfgen} \citep{2007PASJ...59S.113I}.
In {\tt xissimarfgen}, we assumed a uniform sky centered at each observational coordinate with radius of 20 arcmin as an X-ray emitting region for the simulation.
The XDB spectra are contaminated by the NXB which originates from charged particles, electrical noise, and scattered and fluorescent X-ray emission from instrumental elements.
The NXB contributions in the given spectra was also estimated from accumulated night-Earth observations around 150 days centered at the day of each observation with the Ftool {\tt xisnxbgen} \citep{2008PASJ...60S..11T}.
\subsection{Response correction with the Crab Nebula calibration}
The nominal responses deserved above are based on the calibration data and the ray-tracing simulation.
However, a certain level of deviation exists in the response reproduction in the 1.5 -- 3.5 keV due to difficulty of dealing with the complicated multiple absorption edges in the XRT-XIS effective area.
In order to check this reproducibility, we analyzed the stacked spectrum of the Crab Nebula (hereafter simply Crab), the data with the lowest statistical uncertainty of all Suzaku XIS data whose total count of photons in the 0.5 -- 7 keV range is $\sim 4\times 10^7$.
The Crab emission is well modeled by a featureless spectrum of synchrotron emission.
We assumed the emission model for the observed spectrum to be a power-law function absorbed by the interstellar medium of the Milky Way.
We fitted the Crab spectrum with the model multiplied by the response and found the residuals of up to $\sim 10$ \% especially in the 1.5 -- 3.5 keV range.
The deviations were regarded as being due to a mis-reproduction of the response, and these were used to correct the responses.
\subsection{Reproduction of instrumental line contributions}
The NXB including the instrumental emission lines have been estimated and subtracted with night-Earth observational data as described in \citet{2008PASJ...60S..11T}, with a typical error of 3 \%.
There are known instrument lines of Al-K$\alpha$, Si-K$\alpha$, Au-M$\alpha$, Mn-K$\alpha$ and Mn-K$\beta$.
We found that subtleties resulting from the subtraction of these lines can be mitigated by spectral fitting with a five-Gaussian model.
\section{Spectral analysis of the XDB}
\begin{figure}[htbp!]
\begin{center}
   \includegraphics[width=\linewidth]{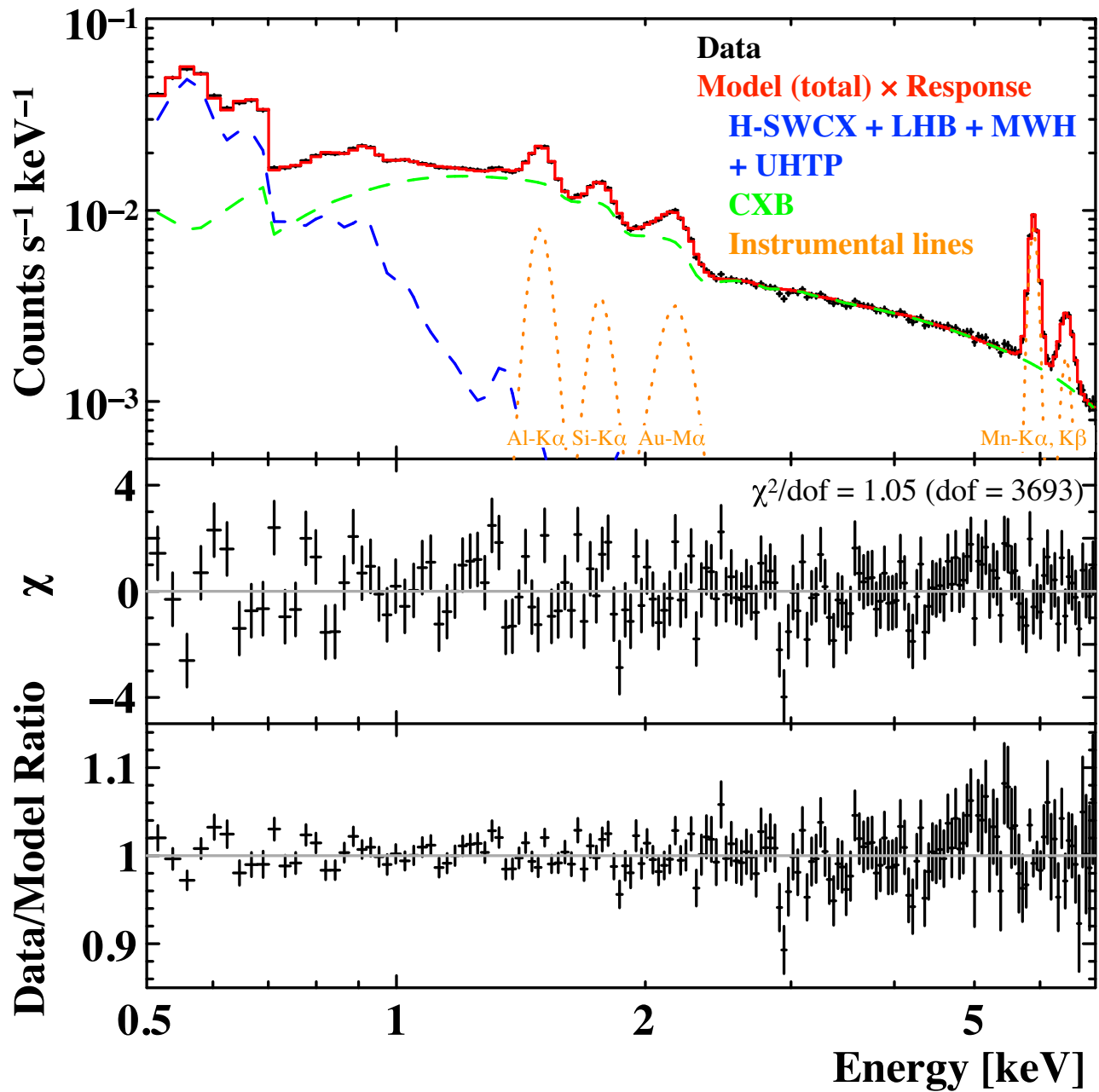} 
\caption{Exposure-time-weighted average of the 25 stacked XDB energy spectra from 2005 to 2013 and its best-fit model convolved with the corrected response.
Sub-components of the model are represented by blue dashed (XDB in the Milky Way (1)+(2)+(4)), green dashed (CXB (3)) and orange dotted (instrumental origin) lines.
}
\label{Figure2}
\end{center}
\end{figure}

Once the 187 sets of the XDB spectra, the RMFs, the ARFs and the NXB data were obtained, we then stacked the XDB spectra using the Ftool {\tt mathpha}.
This was done to increase the number of photons and so to reduce the statistical uncertainty.
With consideration for both the changes in the instrumental-calibration and the difference among the four detectors that make up the XIS, we divided the entire period (2005 -- 2013) into 8 periods to sort by instrumental calibration conditions (e.g. SCI setting) as shown in Table \ref{Table1} and stacked the XDB data in each of the shorter periods.
After correcting for systematic deviations, the total-photon-count-weighted average (in the 0.5 -- 7 keV range) of RMFs $\times$ ARFs (responses) was made by the Ftool {\tt addrmf}.
The exposure-time-weighted average of the NXB data were also stacked.
In total, 25 stacked data sets over eight periods (1 period $\times$ XIS0, 1, 2, 3 + 7 periods $\times$ XIS0, 2, 3) were made.
\begin{table*}[htbp!]
\caption{Stacked data properties.}
\begin{center}
\begin{tabular}{ccccc}
\hline\hline
Period & Date & Total exposure$^\ast$ & Total count$^\dagger$ & Average $N_{\rm H}$$^\ddagger$ \\
\hline
2005--2006 & 2005/10/01 -- 2006/09/30 & 3.2 & 205071 & 0.029 \\
\multicolumn{5}{c}{(SCI operation started for all XISs from October, 2006.)}\\
2006--2007 & 2006/10/01 -- 2007/08/31 & 4.2 & 261725 & 0.035 \\
2007--2008 & 2007/09/01 -- 2008/08/31 & 3.4 & 212512 & 0.029 \\
2008--2009 & 2008/09/01 -- 2009/08/31 & 4.4 & 284447 & 0.030 \\
2009--2010 & 2009/09/01 -- 2010/05/31 & 4.0 & 242187 & 0.030 \\
2010--2011 & 2010/06/01 -- 2011/05/31 & 4.5 & 271709 & 0.029 \\
\multicolumn{5}{c}{(Injection charge increased to 6 keV for XIS1 on June 1, 2011.)}\\
2011--2012 & 2011/06/01 -- 2012/05/31 & 2.3 & 250229 & 0.027 \\
2012--2013 & 2012/06/01 -- 2013/07/01 & 3.5 & 220777 & 0.034 \\
\hline
\end{tabular} 
\begin{flushleft} 
{\bf Notes.}\\
$\ast$ Exposure time (XIS0+1+2+3) in unit of Msec after data screening.\\
$\dagger$ Total photon count in the 0.5 -- 7 keV range.\\
$\ddagger$ The exposure-time-weighted average of the neutral hydrogen column density in unit of $10^{22}$ cm$^{-2}$ derived from the LAB Galactic H\emissiontype{I} Survey.
\end{flushleft}
\label{Table1}
\end{center}
\end{table*}
We then carried out a spectral analysis of the 25 stacked XDB spectra after subtraction of the NXB, using the corrected responses.
We fitted the spectra in the 0.5 -- 7 keV range with the typical XDB emission model of the following components:
\begin{description}
\item[(1)] Heliospheric Solar Wind Charge Exchange (H-SWCX) and Local Hot Bubble (LHB)
\item[(2)] Milky Way hot plasma Halo (MWH)
\item[(3)] Cosmic X-ray Background (CXB)
\item[(4)] Unresolved High Temperature Plasma (UHTP)
\end{description}
(1): the H-SWCX is due to interaction between the Solar wind and neutral atoms in interplanetary space \citep{1998LNP...506..121C,2000ApJ...532L.153C,2006A&A...460..289K,2013PASJ...65...32Y}.
The LHB is considered to be the high temperature plasma with the temperature of $kT\sim 0.1$ keV ($T\sim 10^6$ K) and the density of $n_{\rm H}\sim 0.005$ cm$^{-3}$ embedded in a $\sim 100$ pc cavity of the cold interstellar medium in which the Solar System resides \citep{1982ApJ...253..268C,1990ARA&A..28..657M}.
The H-SWCX and LHB blend is empirically represented by an unabsorbed optically-thin collisionally-ionized (CIE) thermal plasma emission model for the current energy range and the responses of the X-ray CCD cameras.
(2): the MWH is considered to be hot plasma bound in the Milky Way with the temperature of $kT=0.2$ -- 0.4 keV \citep{2009ApJ...690..143Y,2011PASJ...63S.889H,2014PASJ...66...83S}.
It is described by an absorbed optically-thin CIE thermal plasma emission model.
(3): the CXB believed to be superposition of numerous faint extragalactic point sources such as active galactic nuclei and represented by an absorbed power-law model with its photon index $\Gamma\sim 1.4$ \citep{2000Natur.404..459M,2002PASJ...54..327K,2003ApJ...588..696M}.
(4): the UHTP emission sometimes appears in the XDB spectra with strong emission of Fe-L complex and Ne-K lines \citep{2009PASJ...61...805Y,2014efxu.conf...66S}.
It is described by an absorbed optically-thin thermal CIE plasma emission model with the temperature of $kT=0.4$ -- 1.2 keV.
For the optically-thin thermal CIE emission model, we used APEC \citep{2001ApJ...556L..91S,2012ApJ...756..128F}.
The temperature of the APEC model for the H-SWCX + LHB was fixed to $kT=0.1$ keV \citep{2009PASJ...61...805Y}.
The element abundances of the three APEC models were set to the Solar-neighbor values \citep{1989GeCoA..53..197A} except for the Ne and Mg abundances of the UHTP.
The redshifts of the three APEC models were fixed to zero.
In the power-law emission models for the CXB ({\tt powerlaw} in XSPEC), their photon indices were permitted to vary around 1.4.
In Suzaku XIS observational data, an O\emissiontype{I} fluorescent line from the Earth's exosphere was sometimes found especially after 2011, at the time of Solar maximum, despite the fact that its contamination was mostly removed by applying the elevation angle criteria as described above \citep{2014PASJ...66L...3S}.
When we found a discrepancy in the spectral fitting result below 0.6 keV, we added a Gaussian with a centroid of 0.525 keV for O\emissiontype{I} ({\tt gaussian} in XSPEC).
To summarize, the following model for the spectral fitting was adopted: [``APEC$_1$'' + ``Galactic absorption'' $\times$ (``APEC$_2$'' + ``APEC$_3$'' + ``CXB'') + ``O\emissiontype{I}''] where ``Galactic absorption'' was for a photoelectric absorption by the interstellar medium of the Milky Way ({\tt phabs} in XSPEC) which were able to estimate from accurate observational data of the neutral hydrogen column density $N_{\rm H}$ \citep{2005AA...440..775K}, APEC$_1$, APEC$_2$ and APEC$_3$ correspond to the H-SWCX + LHB, the MWH and the UHTP, respectively.
We fitted the stacked XDB spectra with the model for each period.
The results of the XDB spectral fitting and the best-fit parameters of the XDB model are shown in Fig. \ref{Figure2} and Table \ref{Table2}.
As the goodness of fit, $\chi^2/$dof (dof) was 1.05 (3693), and the null hypothesis probability was 2.5 \%.
The deviations between the data and model were about 1.6, 2.9 and 4.0 \% in the root-mean-square at 0.5 -- 2, 2 -- 5 and 5 -- 7 keV, respectively.
\begin{table*}[htbp]
\caption{Spectral fitting results with the stacked energy spectra and the XDB model.}
\begin{center}
\begin{tabular}{ccccccc}
\hline\hline
Period & Norm$_1$$^\ast$ & $kT_2$$^\dagger$ & Norm$_2$$^\ast$ & $kT_3$$^\dagger$ & Ne$^\ddagger$ & Mg$^\ddagger$ \\
\hline
2005--2006 & 20.7$^{+2.4}_{-7.3}$ & 0.22$^{+0.01}_{-0.03}$ & 4.2$^{+1.8}_{-0.4}$ & 0.85$^{+0.06}_{-0.10}$ & 0.0$^{+1.5}_{-0.0}$ & 0.0$^{+1.9}_{-0.0}$ \\
2006--2007 & 22.7$^{+2.5}_{-2.9}$ & 0.24$^{+0.01}_{-0.02}$ & 3.7$^{+0.4}_{-0.4}$ & 0.89$^{+0.10}_{-0.14}$ & 3.2$^{+4.5}_{-3.2}$ & 6.5$^{+3.3}_{-2.9}$ \\
2007--2008 & 19.2$^{+3.7}_{-16.0}$ & 0.23$^{+0.02}_{-0.05}$ & 4.9$^{+4.2}_{-0.6}$ & 0.79$^{+0.10}_{-0.16}$ & 3.8$^{+2.7}_{-2.5}$ & 0.1$^{+2.3}_{-0.1}$ \\
2008--2009 & 20.0$^{+3.5}_{-11.9}$ & 0.22$^{+0.02}_{-0.04}$ & 4.5$^{+3.2}_{-0.5}$ & 0.77$^{+0.09}_{-0.11}$ & 4.6$^{+2.6}_{-2.2}$ & 0.0$^{+1.2}_{-0.0}$ \\
2009--2010 & 14.9$^{+11.3}_{-14.9}$ & 0.19$^{+0.04}_{-0.03}$ & 6.8$^{+5.9}_{-2.7}$ & 0.67$^{+0.10}_{-0.08}$ & 2.9$^{+2.0}_{-1.6}$ & 0.5$^{+1.5}_{-0.5}$ \\
2010--2011 & 42.2$^{+4.6}_{-8.4}$ & 0.24$^{+0.02}_{-0.04}$ & 4.7$^{+1.7}_{-0.6}$ & 0.69$^{+0.08}_{-0.07}$ & 1.6$^{+1.6}_{-1.6}$ & 0.0$^{+0.0}_{-0.0}$ \\
2011--2012 & 34.9$^{+5.6}_{-14.6}$ & 0.22$^{+0.02}_{-0.04}$ & 5.8$^{+3.6}_{-0.7}$ & 0.64$^{+0.14}_{-0.07}$ & 2.7$^{+3.2}_{-1.1}$ & 2.8$^{+2.7}_{-2.0}$ \\
2012--2013 & 36.8$^{+6.6}_{-14.0}$ & 0.22$^{+0.02}_{-0.03}$ & 8.9$^{+3.8}_{-0.8}$ & 0.55$^{+0.07}_{-0.11}$ & 3.3$^{+3.3}_{-1.1}$ & 4.0$^{+3.8}_{-1.3}$ \\
\hline
Period & Norm$_3$$^\ast$ & $\Gamma_{\rm CXB}$$^\S$ & $S_{\rm CXB}$$^\|$ & O\scriptsize{I}$^\sharp$ & \multicolumn{2}{c}{$\chi^2$/dof (dof)} \\
\hline
2005--2006 & 0.4$^{+0.1}_{-0.1}$ & 1.44$^{+0.02}_{-0.02}$ & 7.0$^{+0.2}_{-0.2}$ & 0.4$^{+0.2}_{-0.2}$ & \multicolumn{2}{c}{1.19 (590)} \\
2006--2007 & 0.5$^{+0.1}_{-0.1}$ & 1.42$^{+0.02}_{-0.02}$ & 7.8$^{+0.2}_{-0.2}$ & 0.6$^{+0.2}_{-0.2}$ & \multicolumn{2}{c}{1.41 (434)} \\
2007--2008 & 0.6$^{+0.2}_{-0.2}$ & 1.42$^{+0.02}_{-0.03}$ & 7.8$^{+0.2}_{-0.2}$ & 0.5$^{+0.2}_{-0.2}$ & \multicolumn{2}{c}{1.22 (434)} \\
2008--2009 & 0.6$^{+0.2}_{-0.1}$ & 1.39$^{+0.02}_{-0.02}$ & 7.7$^{+0.2}_{-0.2}$ & 0.4$^{+0.2}_{-0.2}$ & \multicolumn{2}{c}{1.42 (434)} \\
2009--2010 & 0.7$^{+0.3}_{-0.2}$ & 1.48$^{+0.02}_{-0.03}$ & 8.4$^{+0.2}_{-0.2}$ & 0.2$^{+0.3}_{-0.2}$ & \multicolumn{2}{c}{1.26 (434)} \\
2010--2011 & 0.7$^{+0.3}_{-0.2}$ & 1.48$^{+0.02}_{-0.02}$ & 7.9$^{+0.2}_{-0.2}$ & 0.2$^{+0.3}_{-0.2}$ & \multicolumn{2}{c}{1.44 (434)} \\
2011--2012 & 0.6$^{+0.3}_{-0.3}$ & 1.50$^{+0.02}_{-0.02}$ & 8.2$^{+0.2}_{-0.2}$ & 4.2$^{+0.3}_{-0.3}$ & \multicolumn{2}{c}{1.19 (434)} \\
2012--2013 & 0.8$^{+0.4}_{-0.4}$ & 1.35$^{+0.02}_{-0.02}$ & 7.3$^{+0.2}_{-0.2}$ & 5.3$^{+0.4}_{-0.4}$ & \multicolumn{2}{c}{0.99 (434)} \\
\hline
\end{tabular} 
\begin{flushleft} 
{\bf Notes.}\\
$\ast$ The emission measure of the optically-thin thermal CIE plasma integrated over the line of sight (the APEC model normalization):
$(1/2\pi)\int n_{\rm e} n_{\rm H} ds$ in unit of $10^{14}$ cm$^{-5}$ sr$^{-1}$, where $n_{\rm e}$ and $n_{\rm H}$ are the electron and the hydrogen density (cm$^{-3}$).\\
$\dagger$ The temperature of the optically-thin thermal CIE plasma in unit of keV.\\
$\ddagger$ The abundances of Ne or Mg in unit of the Solar-neighbor values given in \citet{1989GeCoA..53..197A}.\\
$\S$ The photon index of the power-law model for the CXB component.\\
$\|$ The surface brightness of the CXB component (the power-law model normalization) in unit of photons cm$^{-2}$ s$^{-1}$ sr$^{-1}$ keV$^{-1}$ at 1 keV (the photon index is fixed at 1.4).\\
$\sharp$ The intensity of neutral oxygen (O\emissiontype{I}, centroid: 0.525 keV) in unit of LU (photons cm$^{-2}$ s$^{-1}$ sr$^{-1}$).
\end{flushleft}
\label{Table2}
\end{center}
\end{table*}
\section{Search for a keV signature of DM}
\begin{figure}[htbp!]
\begin{center}
   \includegraphics[width=\linewidth]{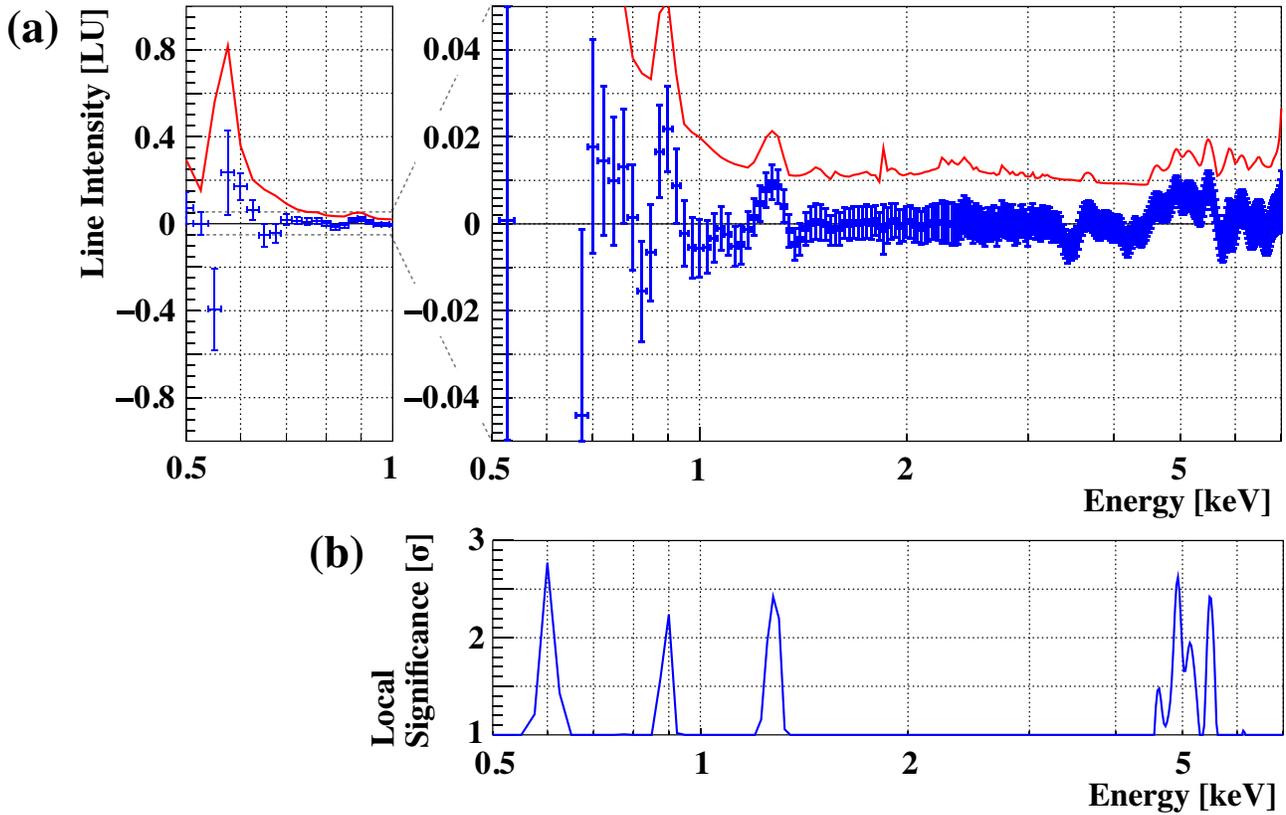} 
\caption{(a) Expected DM line intensity, its 1$\sigma$ statistical error range (blue crosses) and its 3$\sigma$ upper limit (red line).
The unit of line intensity is defined as ``LU'', which is equal to photons cm$^{-2}$ s$^{-1}$ sr$^{-1}$.
(b) Five possible signatures and their statistical significances.
Note that these significances are not and should be corrected by considering the LEE.
}
\label{Figure3}
\end{center}
\end{figure}

\begin{figure}[htbp!]
\begin{center}
   \includegraphics[width=\linewidth]{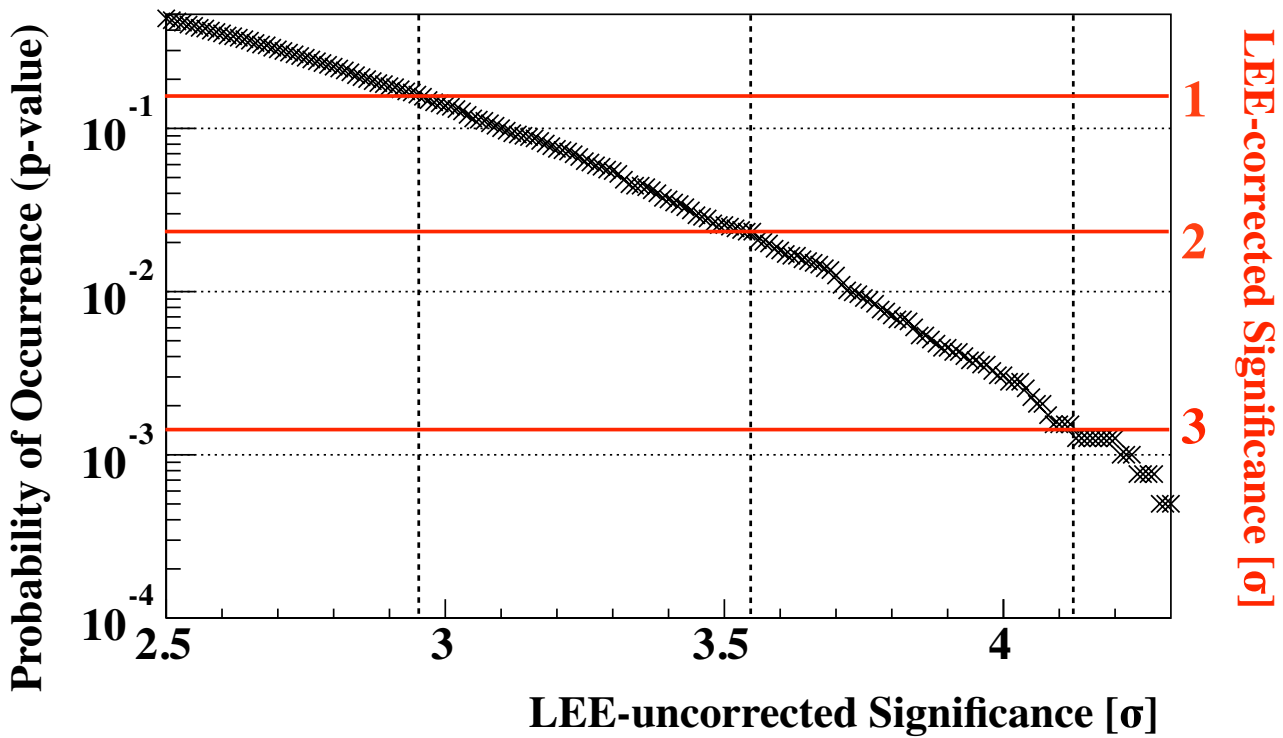} 
\caption{Probability distribution of mis-detection by statistical fluctuation with the 4000 simulations.
The abscissa is the maximum LEE-uncorrected significance for a dummy emission line by random noise in a simulated spectrum, and the ordinate shows the number of occurrence in the 4000 simulated spectra, which is converted into the probability of occurrence (left scale) and the sigma value (right scale; LEE-corrected significance).
The top 0.14, 2.3 and 16 \% of all (corresponding to the one-sided tail of p-value for 3, 2 and 1$\sigma$) are distributed over 4.1, 3.6 and 3.0$\sigma$, respectively.
}
\label{Figure4}
\end{center}
\end{figure}

We simultaneously fitted the stacked XDB spectra with the XDB model with a Gaussian emission line component corresponding to a DM signature.
The line intensity of all the spectra were linked together and allowed to vary freely.
The intrinsic line width of the Gaussian was assumed to be 0 eV because the detection of velocity dispersion due to Galactic rotation or intrinsic to WDM less than several 100 km s$^{-1}$ are impossible with the current instrument.
Its center energy was fixed and was allowed to vary in 25 eV steps over the 0.5 -- 7 keV range.
There were 261 trial lines, and these were not completely independent because the line profiles were broadened by the XIS energy response and these steps were smaller than the energy resolution in the 0.5 -- 7 keV range.
We determined the Gaussian normalization, i.e. the line intensity and its 1$\sigma$ statistical error range.
The results are summarized in Fig. \ref{Figure3} (a) by blue crosses.
The unit of the line intensity is shown by ``Line Unit'' (LU) defined as photons cm$^{-2}$ s$^{-1}$ sr$^{-1}$.
The 3$\sigma$ upper limit on the DM line intensity was calculated as a sum of the best-fit line intensity and its 3$\sigma$ statistical error range, or the only 3$\sigma$ statistical error range when the best-fit line intensity was negative value, i.e. 3$\sigma$ upper limit = max$\{$best-fit line intensity, 0$\}$ + 3$\sigma$ statistical error range.
This upper limit was shown in Fig. \ref{Figure3} (a) by the red line.
We found line-like signatures at 0.600, 0.900, 1.275, 4.925 and 5.475 keV as shows in Fig. \ref{Figure3} (b), with 2 -- 3$\sigma$ statistical significances.
Since the energy of a DM emission line is not known {\it a priori}, it is necessary to take into account the number of independent line searches per one spectral fitting, i.e. the ``trial factor''.
For example, if we have a trial factor of 100 in a spectrum, the expectation to detect a line from random noise with an intensity exceeding the 99 \% significance is unity.
This effect is generally considered in the pulsation search from timing data (e.g. \cite{1991ApJ...379..295W}), and is also sometimes called ``Look-elsewhere Effect'' (LEE; \cite{2010EPJC...70..525G}).
The trial factor is determined by the energy range and the energy resolution.
This is not analytically known, since the energy spread function is relatively broad and since the energy resolution is energy-dependent.
We therefore conducted Monte-Carlo simulations.
We first generated a simulation spectrum assuming the best-fit XDB model without an additional emission line.
We then determined the maximum significance for a dummy emission line in each simulated spectrum.
This procedure was iterated 4000 times, and the cumulative occurrence distribution of the 4000 maximum significances was plotted in Fig. \ref{Figure4}.
From this figure, we found that significances $\ge 3.0$$\sigma$ appeared 635 times in the 4000 simulations (i.e. 16 \%; the one-sided tail of p-value for 1$\sigma$).
This means that LEE-uncorrected 3.0$\sigma$ corresponds to LEE-corrected 1$\sigma$.
Similarly, we found that LEE-uncorrected 4.1$\sigma$ corresponded to LEE-corrected 3$\sigma$.
As a result, line-like signatures found in this search should be regarded as statistical fluctuations.
We also evaluated the systematic uncertainty in this line search.
Three main causes were investigated:
\begin{enumerate}
\item XDB model uncertainty.
\item Instrumental response uncertainty.
\item NXB contribution uncertainty.
\end{enumerate}
The XDB model especially below 1 keV contains multiple emission lines from the hot plasma, i.e. the SWCX+LHB blend, the MWH and the UHTP, and their element abundances are difficult to determine.
We varied their element abundances from 1/2 to $2\times$ the Solar-neighbor values and checked the effects on the line search results.
The instrumental responses between 1.5 and 3.5 keV were corrected by the Crab calibration.
The statistical uncertainty of this correction were also considered to give systematic errors.
For the NXB reconstruction, the errors of the five instrumental lines were evaluated as the standard deviation of those obtained by the night-Earth data corresponds to the 187 blank sky data used in this analysis.
\section{Discussion}
\begin{figure}[htbp!]
\begin{center}
   \includegraphics[width=\linewidth]{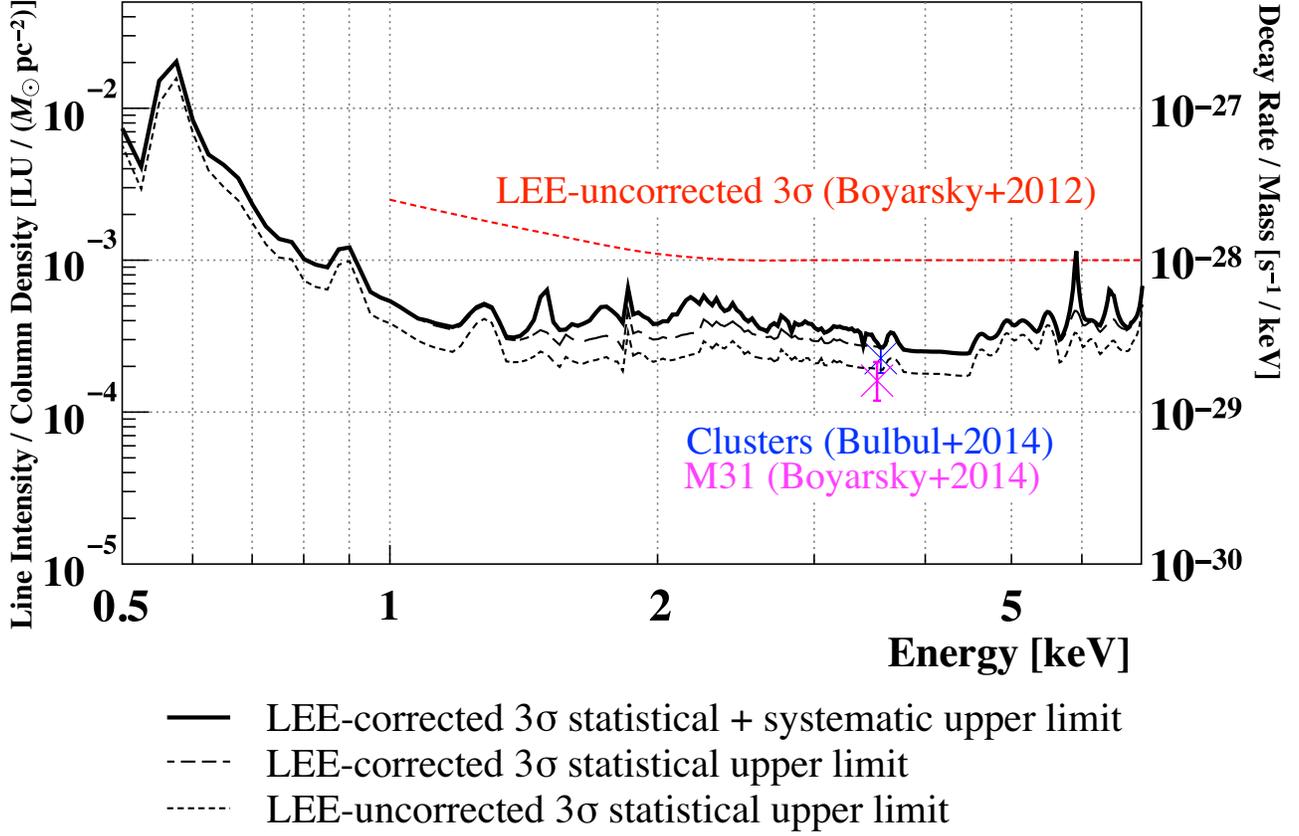} 
\caption{Upper limit on the DM line intensity normalized by its column density corresponding to the ratio of DM decay rate and its mass.
The LEE-uncorrected, -corrected 3$\sigma$ and the LEE-corrected 3$\sigma$ statistical + systematic upper limits are indicated by the black dotted, dashed and solid lines, respectively.
The typical 3$\sigma$ upper limit by previous works \citep{2012PDU.....1..136B} is expressed by the red dotted line.
The parameters of the possible 3.5 keV line \citep{2014ApJ...789...13B,2014PhRvL.113y1301B} are plotted by the blue and magenta crosses.
}
\label{Figure5}
\end{center}
\end{figure}

\begin{figure}[htbp!]
\begin{center}
   \includegraphics[width=\linewidth]{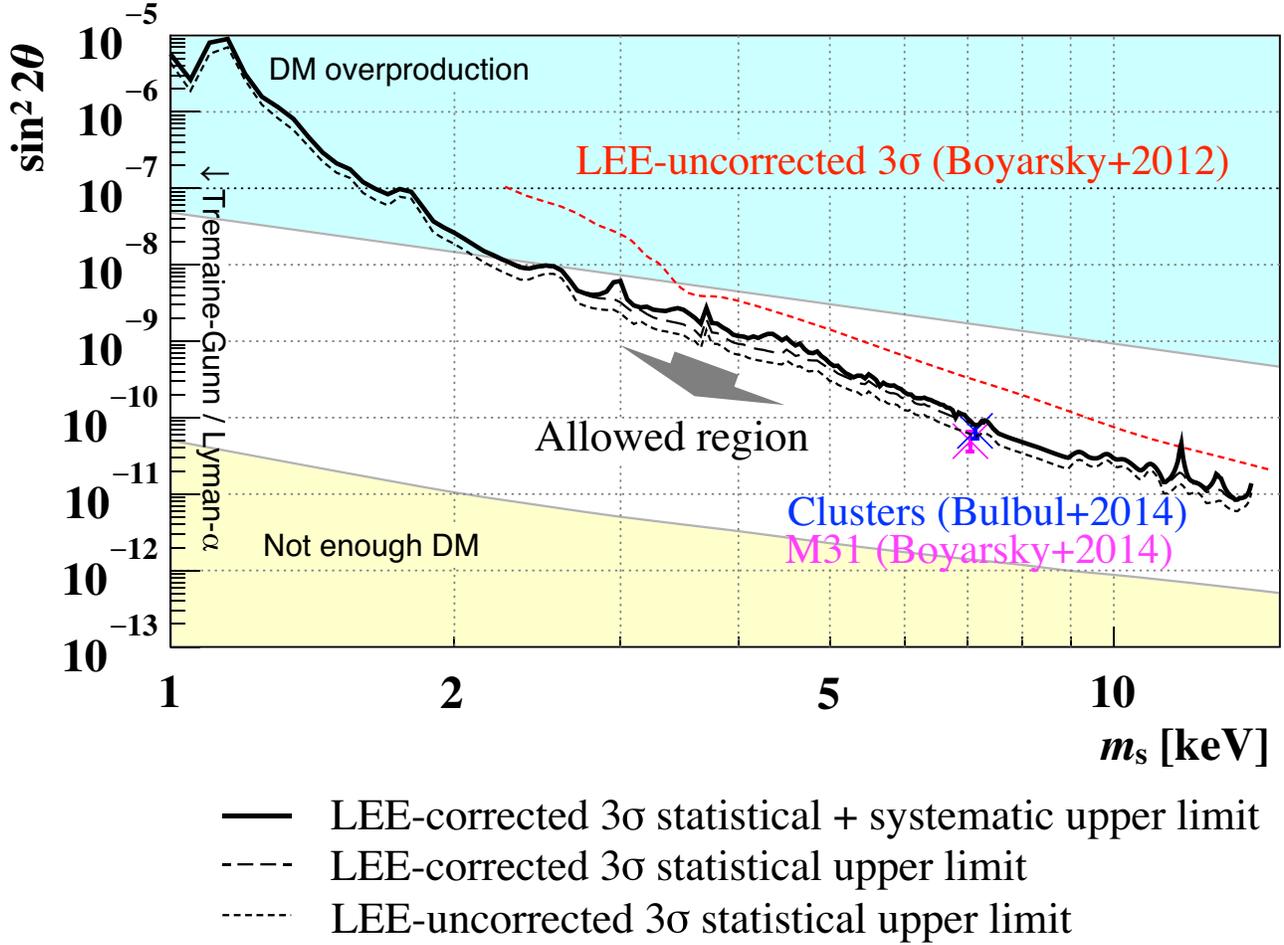} 
\caption{Constraints on the sterile neutrino mass $m_{\rm s}$ and mixing angle $\sin^2 2\theta$ by this and previous works.
The definition of lines and marks are the same as in Fig. \ref{Figure5}.
The cyan and yellow shaded regions are excluded by the production theories of sterile neutrinos in the $\nu$MSM (see text for details).
}
\label{Figure6}
\end{center}
\end{figure}

No significant signature of radiatively decaying DM was found in the 0.5 -- 7 keV range in this search.
This result is consistent with the stacked spectral analysis of the Perseus cluster by using Suzaku XIS observational data \citep{2015PASJ...67...23T}.
The DM line intensity $I$ is expressed by the DM column density $S_{\rm DM}$, the decay rate $\Gamma$ and the mass $m_{\rm DM}$:
\begin{equation}
I=\frac{S_{\rm DM} \Gamma}{4\pi m_{\rm DM}}.\label{Eq1}
\end{equation}
We are able to convert the DM line intensity normalized by its column density into its decay rate divided by its mass.
We obtained the upper limit on the ratio of the DM decay rate and its mass as shown in Fig. \ref{Figure5} in comparison with previous works \citep{2012PDU.....1..136B,2014ApJ...789...13B,2014PhRvL.113y1301B}.
The exposure-time-weighted average of DM column density in this work was 63 $M_\odot$ pc$^{-3}$ from the NFW model \citep{2012PASJ...64...75S}.
The LEE-uncorrected upper limit was also plotted in addition to the LEE-corrected one because the previous works have not adapted the LEE-correction, and for the 3.5 keV upper limit, it does not need to be considered.
Possible systematic uncertainties were also included.
Assuming sterile neutrinos to make up DM, we converted this observational result into constraints on their mass and mixing angle with the active neutrinos as the same manner in \citet{2001ApJ...562..593A}.
The decay rate of sterile neutrinos is written as
\begin{equation}
\Gamma=\frac{9\alpha {G_{\rm F}}^2}{1024\pi^2}{m_{\rm s}}^5 \sin^2 2\theta
=1.4\times10^{-32}\ {\rm s^{-1}}\ \left(\frac{m_{\rm s}}{1\ {\rm keV}}\right)^5 \left(\frac{\sin^2 2\theta}{10^{-10}}\right),\label{Eq2}
\end{equation}
where $\alpha$ is the fine-structure constant, $G_{\rm F}$ is the Fermi constant and $\theta$ is a sterile neutrino mixing angle \citep{1982PhRvD..25..766P}.
From Eq. \ref{Eq1} and \ref{Eq2}, the line intensity is
\begin{equation}
I=1.3\times10^{-5}\ {\rm LU}\ \left(\frac{m_{\rm s}}{1\ {\rm keV}}\right)^4\left(\frac{\sin^22\theta}{10^{-10}}\right)\left(\frac{f_{\rm s}}{1}\right)\left(\frac{S_{\rm DM}}{10^2 M_\odot {\rm pc}^{-2}}\right),\label{Eq3}
\end{equation}
where $f_{\rm s}$ is the fraction of $\nu_{\rm s}$ in DM.
From Eq. \ref{Eq3}, we obtained the constraint on their mass and mixing angle as shown in Fig. \ref{Figure6}.
The cyan and yellow shaded regions have been already rejected by several studies, such as the non-resonant (cyan shaded region; \cite{2009JCAP...03..005B}), the resonant production with the maximal lepton asymmetry attainable in the $\nu$MSM (yellow shaded region; \cite{2008JHEP...08..008S,2008JCAP...06..031L}).
The Tremaine-Gunn phase-space density considerations \citep{2009JCAP...03..005B} and the Lyman-$\alpha$ analysis \citep{2009JCAP...05..012B,2009PhRvL.102t1304B} ruled out the region below 1 keV.
We have added an exclude region for $\sin^2 2\theta$ in $3 < m_{\rm s} < 14$ keV.
\citet{2014PhRvD..89b5017H} also claimed a tight upper limit by using Chandra ACIS observational data of M31 with exposure time of $\sim400$ ksec.
We did not include it in our plot because their confidence level definition with fit statistics was not clear.
\section{Conclusion and future prospects}
In this paper, we searched for the signature of an X-ray emission line from DM associated with the Milky Way halo by using a set of 187 Suzaku XIS archival data sets of the XDB from 2005 to 2013.
After improving reproducibility of the instrumental responses and the instrumental line contributions, we searched for a non-baryonic emission line in the stacked XDB spectra by spectral fitting with $[$(the corrected response by using the stacked Crab spectra) $\times$ (the best-fit XDB model + five-instrumental lines + Gaussian emission line model for a DM signature)$]$.
Consequently, we did not detect a possible DM signature including the line at 3.5 keV reported by previous studies (e.g. \cite{2014ApJ...789...13B}), and determined the upper limit on the emission line intensity taking into account the LEE.
We tightened the constraints on the ratio of DM decay rate and its mass in the 0.5 -- 7 keV range and the parameters of sterile neutrinos.
In the future, progressive instruments such as X-ray micro-calorimeters with energy resolution for diffuse X-ray emission 
of an order of eV and the large-FoV telescopes will be introduced to X-ray observational satellites, and more sensitive DM searches will be performed.
Among the more hopeful instruments of near future missions, we are focussed on the Soft X-ray Spectrometer (SXS) of the ASTRO-H satellite \citep{2010SPIE.7732E..0ZT,2014SPIE.9144E..2AM} and the extended ROentgen Survey with an Imaging Telescope Array telescope with the PN-CCD camera module (eROSITA) of the Spektrum-Roentgen-Gamma (SRG) satellite \citep{2014SPIE.9144E..1TP,2014SPIE.9144E..1WM}.
The ASTRO-H SXS is an X-ray micro-calorimeter with doped semiconductor thermistors and will have the highest energy resolution, other than for grating instruments for point sources, though its grasp is lower than that of the existing X-ray observatories.
On the other hand, the SRG eROSITA has the largest grasp and an all sky survey plan which is suitable for deeper analysis of the XDB, though its energy resolution is more modest.
Especially in the ASTRO-H SXS with the high line identification ability by the high energy resolution, it is suitable for a weak line search with ``dense'' targets such as clusters of galaxies and nearby galaxies although their background plasma emission are strong.
The future X-ray observations will give a tighter constraint on DM conditions.
\begin{ack}
This work is partly supported by JSPS KAKENHI Grant Numbers 12J10673, 25247028 and 26220703.
We thank Dr. K. Sakai for valuable suggestions.
\end{ack}

\end{document}